# METIS – the Mid-infrared E-ELT Imager and Spectrograph

Bernhard R. Brandl[*a], Rainer Lenzen[b], Eric Pantin[c], Alistair Glasse[d], Joris Blommaert[e], Lars Venema[f], Frank Molster[a], Ralf Siebenmorgen[g], Hermann Boehnhardt[h], Ewine van Dishoeck[a], Paul van der Werf[a], Thomas Henning[b], Wolfgang Brandner[b], Pierre-Olivier Lagage[c], Toby J.T. Moore[i], Maarten Baes[k], Christoffel Waelkens[e], Chris Wright[l], Hans Ulrich Käufl[g], Sarah Kendrew[a], Remko Stuik[a] & Laurent Jolissaint[a]

[a]Leiden Observatory, Leiden University, P.O. Box 9513, 2300 RA Leiden, The Netherlands;
[b]Max-Planck-Institut für Astronomie, Königstuhl 17, 69117 Heidelberg, Germany;
[c]Groupe LFEPS, Service d'Astrophysique, CE Saclay DSM/DAPNIA/Sap, 91191 Gif sur Yvette Cedex, France;
[d]UK Astronomy Technology Centre, Edinburgh EH9 3HJ, UK;
[e]Instituut voor Sterrenkunde, K.U.Leuven, Celestijnenlaan 200D, B-3001 Leuven, Belgium
[f]ASTRON, P.O.Box 2, 7990 AA Dwingeloo, The Netherlands;
[g]European Southern Observatory, Karl-Schwarzschild-Strasse 2, D-85748 Garching, Germany;
[h]Max-Planck-Institut für Sonnensystemforschung, Max-Planck-Str. 2, 37191 Katlenburg-Lindau, Germany;
[i]Astrophysics Research Institute, Liverpool John Moores University Twelve Quays House, Egerton Wharf, Birkenhead CH41 1LD, UK;
[k]Sterrenkundig Observatorium, Universiteit Gent, Krijgslaan 281 S9, B-9000 Gent, Belgium;
[l]School of Physical, Environmental and Mathematical Sciences, University of New South Wales, Canberra, Australia.

**ABSTRACT**

METIS, the Mid-infrared ELT Imager and Spectrograph (formerly called MIDIR), is a proposed instrument for the European Extremely Large Telescope (E-ELT), currently undergoing a phase-A study. The study is carried out within the framework of the ESO-sponsored E-ELT instrumentation studies. METIS will be designed to cover the E-ELT science needs at wavelengths longward of 3μm, where the thermal background requires different operating schemes. In this paper we discuss the main science drivers from which the instrument baseline has been derived. Specific emphasis has been given to observations that require very high spatial and spectral resolution, which can only be achieved with a ground-based ELT. We also discuss the challenging aspects of background suppression techniques, adaptive optics in the mid-IR, and telescope site considerations. The METIS instrument baseline includes imaging and spectroscopy at the atmospheric L, M, and N bands with a possible extension to Q band imaging. Both coronagraphy and polarimetry are also being considered. However, we note that the concept is still not yet fully consolidated. The METIS studies are being performed by an international consortium with institutes from the Netherlands, Germany, France, United Kingdom, and Belgium.

## 1. INTRODUCTION

METIS, the '**M**id-infrared **E**LT **I**mager and **S**pectrograph', is named after the first spouse of Zeus and mother of Athena, the goddess of wisdom in Greek mythology. The main philosophy behind the METIS concept is to cover the mid-infrared wavelength range longward of 3μm[1] with one dedicated instrument. The instrument has been selected by ESO for a conceptual phase-A study as a potential first-generation E-ELT instrument. (The concept for METIS has already been studied within an EU-funded "Small Study" and has been selected for a follow-up point-design study, which started in November 2007.) The METIS consortium consists of institutions from five ESO countries: the Netherlands (NOVA, Astron), Germany (MPIA), France (CEA Saclay), the United Kingdom (UK-ATC), and Belgium (KU Leuven). The METIS team draws heavily from its experience with numerous successful ground- and space-based instruments, such as VLT-VISIR, VLTI-MIDI, GEMINI-MICHELLE, Spitzer-IRS and JWST-MIRI.

---
[*] brandl@strw.leidenuniv.nl; phone +31 (0)71 527 5830

The great success of the Spitzer Space Telescope, and the high expectations in the James Webb Space Telescope, indicate the great scientific potential of observations at mid-IR wavelengths. Unfortunately, mid-IR astronomy from the ground has a bad reputation of not being competitive to observations from space. The very high background on the ground will undoubtedly lead to reduced sensitivity, in particular to low surface brightness features, and less stable conditions. Furthermore, atmospheric absorption limits the wavelength coverage to specific windows. However, the huge collecting area of the E-ELT and the resulting unsurpassed spatial and spectral resolving power will provide many advantages over space, providing access to complementary parts of the parameter space. The three most relevant advantages on the ground are:

- Much higher angular resolution due to the larger aperture. In comparison with JWST the E-ELT will deliver 6.5 times sharper images and spatially resolved spectroscopy.

- Very high spectral resolution for studies of narrow spectral features. High resolution spectrographs have sizes and weights that are beyond what can be fitted into a typical science payload module of a space mission.

- Flexibility and higher complexity due to almost real-time access at little cost. That enables the implementation of more instrument modes (e.g. polarimetry), using the latest technologies available, and provides an upgrade path for the instrument.

## 2. THE METIS SCIENCE CASE

The identification of the main science drivers for METIS is based on two fundamental considerations:

(1) What are the scientifically most important observations ("Killer science") that can be best done in this wavelength range?

(2) What areas in parameter space will be best or uniquely covered by the E-ELT, in particular in comparison to observations in space.

The METIS science case[2] is closely linked to the *ASTRONET Science Vision for European Astronomy*, and will make significant contributions to numerous areas from the "Extremes of the Universe" to "How … we fit in". The *ASTRONET Science Vision* concluded on page 138 that "(...) mid-infrared imaging and spectroscopy at high spatial resolution and sensitivity provided by an Extremely Large Telescope with high performance adaptive optics will be essential". METIS will also be able to make significant contributions to five of the nine science drivers for the E-ELT as defined by ESO's Science Working Group: 'Exo-planets', 'Stellar disks', 'IMF of clusters + Galactic Center', 'Black holes /AGN', and 'The first light'.

In the following subsections we summarize the four main science drivers; more details can be found in Brandl et al. (2008).

---

[1] Here we use the term "mid-infrared" to generally refer to wavelengths $\lambda \geq 3\mu m$.
[2] The METIS science case has been assembled by the members of the METIS science team (who co-author this paper) with valuable contributions from Geoff Blake (Caltech), Sebastian Dämgen (MPIA), Kerstin Geißler (MPIA), Miwa Goto (MPIA), Markus Janson (MPIA), Lex Kaper (Amsterdam), Fred Lahuis (SRON/Leiden), Hendrik Linz (MPIA), Klaus Pontoppidan (Caltech), Colette Salyk (Caltech), Michael Sterzik (ESO), Sebastian Wolf (Kiel), and Klaas Wiersema (Amsterdam).

## 2.1 The Solar System

The bodies in the planetary system were formed 4.6 billion years ago in an extended disk, and the formation took only a few ten million years. The physical conditions and the chemical composition of the planetary formation disk at that time is hardly constrained except that strong radial gradients in temperature and density are suggested from the inner edge close to the Sun, where the terrestrial planets originated, to the region of the gas giants, where the icy bodies like comets and Kuiper Belt objects accreted. The physical processing of that formation disk material is, at least partially, reflected in properties of the most primordial bodies in the solar system that are accessible to observations today: comets and other minor bodies. Of particular interest is how they connect to Earth, in particular to the existence of water and the formation of life. Their physical properties provide a direct link to their origin in the formation disk and to the physical conditions and the chemical environment therein. While comets are representing the outer formation disk, where the giant planets Jupiter, Saturn, Uranus and Neptune were formed, asteroids are characteristic of the region where terrestrial planets form. It is thus of outmost scientific interest to explore these relics from the planetesimal era of the protoplanetary disk. This will lead to a more comprehensive understanding of the conditions and the processes that have led to the formation of our own planetary system and that are still at work in formation disks around other stars in our galaxy.

Since comets were formed at diverse distances from the Sun and since they reside most of their lifetime in the cold environment beyond and even much beyond the planetary frost line at around 3-4 AU from the Sun, cometary nuclei are believed to contain the most primordial material in our solar system that is accessible to Earth-based observations. The scientific goals for the exploration of the era of planetesimal formation in our planetary system can be accomplished by:
1. establishing a composition and temperature profile of the planetary formation disk
2. determining isotope ratios in cometary volatiles, namely the D/H ratio of cometary water and how it relates to that of terrestrial $H_2O$ as well as the ratios of $^{12}C/^{13}C$ and $^{14}N/^{15}N$, important elements in organic compounds
3. constraining the large-scale radial mixing of material in the disk
4. estimating thermal inertia of asteroidal bodies and cometary nuclei to constrain their internal constitution.

### 2.1.1 Composition, temperature profile and isotopic ratios in the formation disk

The ices in comet are expected to reflect best the gaseous composition in the formation disk of the Sun by the time shortly before the planets were formed. Sublimation of the ices when the comet gets closer to the Sun, releases volatile species. The richest wavelength domain for studies of these ice volatiles is the IR region of 3 – 5μm (Mumma et al. 2003). Using high-dispersion spectroscopy, a number of parent gas species from cometary ices – including organic compounds – can be measured, for instance $H_2O$, CO, $NH_3$, $CH_4$, $C_2H_2$, $C_2H_6$, $CH_3OH$, HCN. A number of emission lines from so far unidentified species is also seen in high dispersion 3 – 5μm spectra of bright comets. Considering evolutionary effects of cometary nuclei, the production rates of the known molecules allow to conclude on the ice composition in the outer planetary formation disk. A good indicator of the temperature domain in the region where the comets were formed is the ortho-to-para ratio of parent molecules like $H_2O$ and $NH_3$ (dello Russo et al. 2005). It allows determining the spin temperature of these species, which is considered a good proxy for the formation temperature of the respective ices. Temperature models of the formation disk can then provide the link to the likely formation distance of the comet in the proto-planetary disk. The radial profiles of both the gas composition and the temperature regime in the disk evolve from measurements of a representative sample of short- and long-periodic comets. Here, over the past ten years, a starting point was the observations of a few, mostly long-periodic comets from the Oort Cloud. However, progress is slow since the objects must be very bright and thus rare. In particular, short-periodic comets are not well represented in the small sample of objects which can be measured with today's telescopes.

METIS will also be suitable to assess the isotopic composition of cometary ices using mostly the same wavelength regions as for normal gas production rate estimations. Here, the D/H ratio in cometary water is of particular interest since it is suspected that terrestrial water may – at least partially – come from cometary impacts on the early Earth, for instance during the late heavy bombardment by Kuiper Belt objects about 600 million years after formation. Existing D/H measurements (for a review see Bockelee-Morvan et al. 2004) indicate a higher ratio in comets than in terrestrial water; however, they only refer to Oort Cloud comets which may not have participated in the heavy late bombardment of the Earth. Combining isotopic ratio values in comets – not only D/H, but also $^{12}C/^{13}C$ and $^{14}N/^{15}N$ as important elements in organic molecules – with the compositional and temperature radial profile of the formation disk will answer the question of isotopic enrichment and attenuation processes in the disk. The key contribution for the isotopic analysis can only come from measurements of more objects, in particular short-periodic comets are difficult since they require mostly telescope with apertures larger than currently available.

**2.1.2 Large-scale radial mixing of disk material**

The composition mixing of gaseous species in the planetary disk follows from the above mentioned L and M band spectroscopy of gases in cometary comae. The solid, non-volatile component, namely the silicates, can be measured by N band spectroscopy at low resolution. Recently, a mixture of amorphous, space-weathered and crystalline silicates (Hanner & Bradley 2004, Lisse et al. 2005) was found in two long- and one short-periodic comet (Figure , left). The co-existence of amorphous and crystalline phases in cometary dust suggests that part of the crystalline silicates may have been processed in a hot environment above the melting temperature (for a review see Wooden 2003). This happened most likely close to the Sun, just shortly before they got inserted into the cometary nucleus further away from the Sun. This scenario requires efficient radial mixing of material in the proto-planetary disk, a process that was so far not considered important in the formation scenario of our planetary system. At present it is completely unknown to which extent mixing took place and whether or not it affected the whole disk.

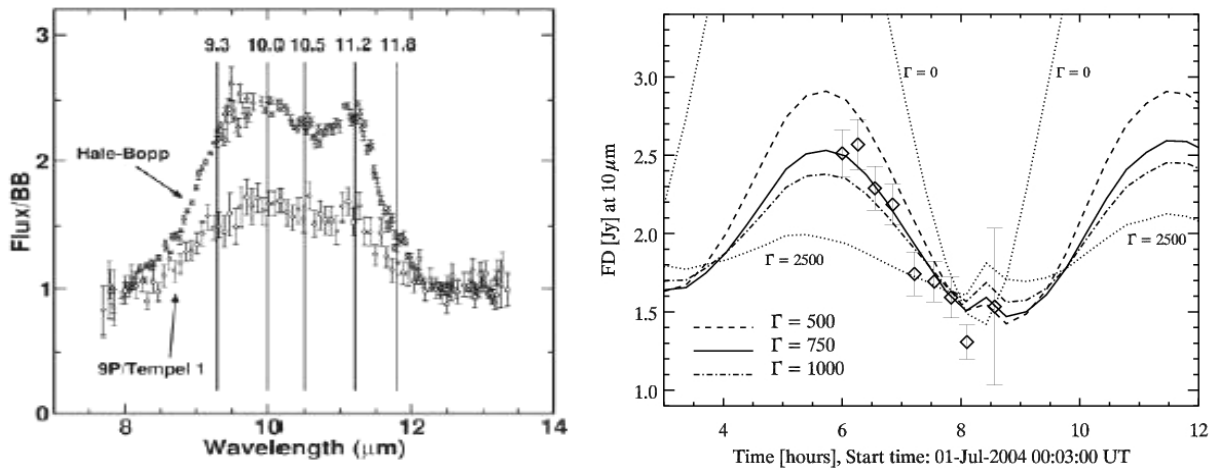

Figure 1: *Left:* N band spectra of comets C/1995 Hale-Bopp and 9P/Tempel 1 (Harker et al. 2005). The peaks at 11.2μm are due to the presence of crystalline silicate grains in the dust coamae of the comets, the emission shortwave thereof is produced by amorphous silicates. *Right:* N band rotation curve (diamond symbols) of asteroid 25143 Itokawa and thermal modelling results for various parameters of thermal inertia $\Gamma$ (Müller et al. 2005).

**2.1.3 Thermal inertia and the internal constitution of minor bodies**

Recent improvements in thermal modeling techniques allow meanwhile to estimate thermal properties of the surfaces of minor bodies from mid-IR photometry. The thermal emission of objects in Earth vicinity peaks at mid-IR wavelengths. Combined N- and Q-band measurements at different rotational phases over a range of phase angles are key ingredients to determine the thermal inertia. The thermal inertia of large, dust-covered main-belt asteroids is found to be very low, while monolithic bedrock objects have thermal inertia two orders of magnitude higher (Harris & Lagerros 2002). Tidal forces during encounters with large bodies in the solar system cause regular reorganizations of such rubble piles (Bottke et al. 2003), resulting in thermal properties which are characteristic for a mixture of dusty and rocky surface regions.

First indications on the internal constitution of a small body came from N-bands observations (Müller et al. 2005) at a 4m-class telescope (Fig. 1, right). The principle capability of constraining the internal structure of minor bodies from the thermal properties allows assessing the constitution of rather primordial bodies like cometary nuclei and C- and D-type asteroids as well as more evolved ones like S- and M-type asteroids. The scenarios for planetesimal formation in the proto-planetary disk, as well as for the impact of the collision history on the body constitution, are expected to evolve and require observational constraints. The key are accurate thermal measurements of minor bodies for which N and Q band photometry on an ELT is required. It is noteworthy that for objects of 10 km and larger, surface resolved measurements will be possible.

## 2.2 Exoplanets

This science case aims at detecting exoplanets around stars in the solar neighbourhood, and at studying the physical and chemical properties of these exoplanets. Since the announcement of the first exoplanet identified around 51 Peg by radial velocity variations (Mayor & Queloz 1995), more than 200 exoplanets have been discovered. Yet the *direct* detection of exoplanets around other stars remains extremely challenging. Mid-infrared studies of exoplanets promise to be particularly rewarding for three reasons: First, as this wavelength region encompasses a plethora of spectral lines and band structures from the multitude of molecules present in planetary atmospheres. Second, the contrast between Jupiter and our Sun is one to 500 million in the visual (V-band) and one to 2 billion in the near-infrared (J-band); the situation improves considerably in the mid-infrared to contrasts of one to 20 million in the M-band and one to 2 million at 11.3 μm (N-band), respectively. Third, observations at longer wavelengths are also accompanied by a better behaved point spread function with the Strehl ratios increasing from approximately 65% in the H-band and 80% in the L-band, to 93% at 11 μm[3].

### 2.2.1 Direct detections via imaging

We have computed a grid of flux densities for various stellar types, distances from the Sun, and star-planet separations[4]. Table 1 summarizes the expected flux from exo-earths, super-earths, and "ice giants" located in the centre of the habitable zone around their parent star. Figure 2 shows simulated observations for two stellar systems of different ages and planet masses. The simulations include the effects of sky shot noise and PSF variations[5] and assume a mid-IR optimized E-ELT site, such as Cerro Macon. For angular separations smaller than about 0.15 arcsec the point source sensitivity is clearly dominated by PSF subtraction errors.

Table 1: Flux densities from a terrestrial planet (1 $R_E$, 2 $R_E$) and an exo-Neptune (4$R_E$) orbiting in the centre of the habitable zone. For an M0V star, this corresponds to a=0.3 A.U., for a G2V star to a=1 A.U., and for an A2V star to a=5 A.U.

|   | $F_{M'}$ [μJy] | | | $F_N$ [μJy] | | | $F_Q$ [μJy] | | |
|---|---|---|---|---|---|---|---|---|---|
| d | 1 $R_{Earth}$ | 2 $R_{Earth}$ | 4 $R_{Earth}$ | 1 $R_{Earth}$ | 2 $R_{Earth}$ | 4 $R_{Earth}$ | 1 $R_{Earth}$ | 2 $R_{Earth}$ | 4 $R_{Earth}$ |
| 1 pc | 4.0 | 16 | 64 | 28 | 110 | 450 | 8.0 | 32 | 128 |
| 2 pc | 1.0 | 4.0 | 16 | 7.0 | 28 | 110 | 2.0 | 8.0 | 32 |
| 3 pc | 0.4 | 1.8 | 7.1 | 2.8 | 13 | 50 | 0.8 | 3.6 | 14 |
| 4 pc | 0.24 | 1.0 | 4.0 | 1.7 | 7.0 | 28 | 0.48 | 2.0 | 8.0 |
| 5 pc | 0.16 | 0.64 | 2.6 | 1.1 | 4.5 | 18 | 0.32 | 1.2 | 5.2 |

Additional PSF suppression using differential techniques, like spectral differential imaging (SDI) or an advanced coronagraph, will aid the direct detection of exoplanets. Fig. 3 (left) shows the limiting sensitivities of METIS for companions as a function of distance from a solar type star at 5 pc. Two exposures have been simulated, with optical seeings of 0.6" (target star) and 0.8" (reference star), respectively. With a four quadrant phase mask (4QPM) coronagraph, the starlight rejection is 3 – 10 times better than standard PSF subtraction alone, enabling the detection of 5$M_J$ planets down to 1 AU from the star. Furthermore, the 4QPM will strongly reduce detector artifacts that would be caused by the bright star.

### 2.2.2 Spectroscopic detections – transiting planets

The atmospheres of giant planets consist primarily of molecular hydrogen and helium, with traces of various other molecules. Examples of prominent thermal IR diagnostic lines, which are accessible from the ground, are $NH_3$ (ammonia) and $PH_3$ (phosphine) at 5-5.4μm, $CH_4$ (methane) in absorption at 3.5 – 4.1μm and emission at 7 – 14μm, $C_2H_2$ at 13.5μm, $C_2H_4$ at 10.4μm, and $H_3^+$ at 3.3μm. The latter contributes some of the strongest and most prolific spectral features in the high-resolution spectrum of Jupiter and is generally expected to be the dominant coolant in the upper atmospheres of giant planets.

---

[3] Simulations performed by ESO's AO group, assuming a wavefront correction by the M4 of the E-ELT.

[4] For details of the calculation see Brandl et al. (2008).

[5] To simulate PSF variations, AO corrected PSFs were computed for seeing values of 0.75" and 0.70" and subtracted.

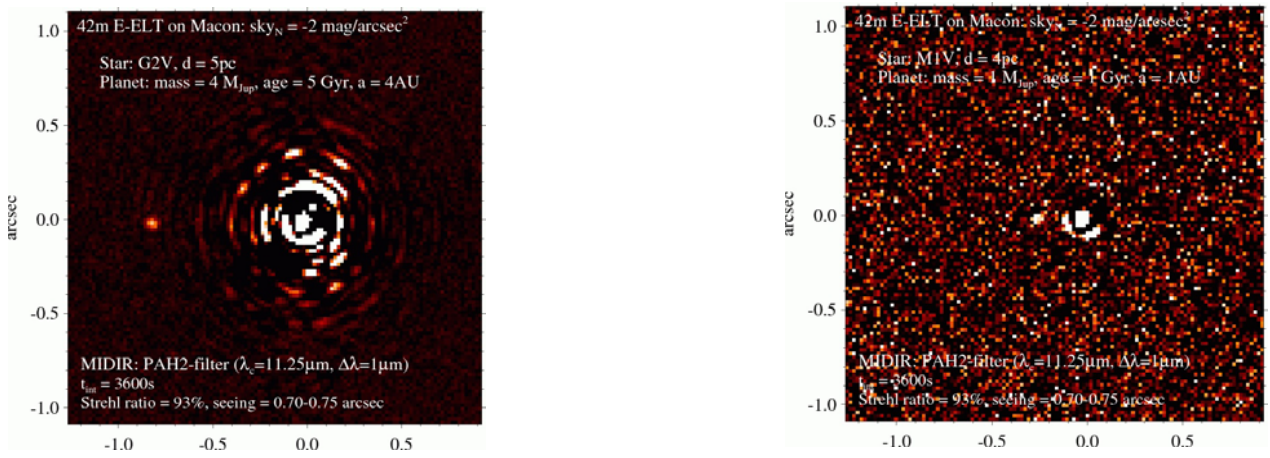

Figure 2: *Left:* Simulated METIS observations of a 5 Gyr old, 4 $M_J$ giant exoplanet at 4 A.U. (0.8" at 5 pc) from a G2V star as seen through a PAH 11.25μm filter. *Right:* Simulated METIS observations of a 1 Gyr old, 1 $M_J$ giant exoplanet at 1 AU from a M1V star at a distance of 4 pc from the Sun. Both simulations have been made for a good mid-IR site (Cerro Macon).

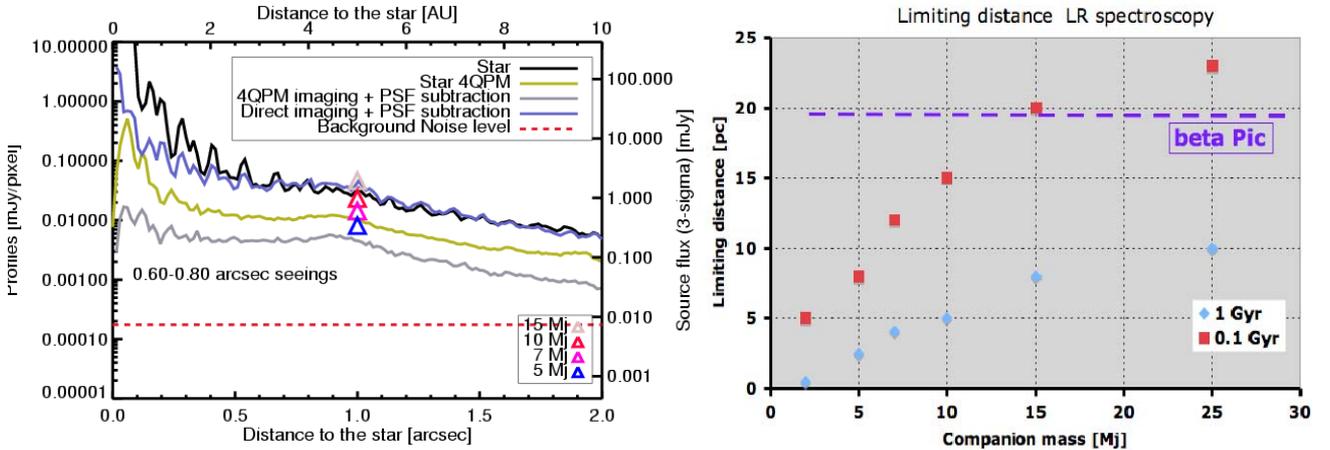

Figure 3: *Left:* Limiting METIS sensitivities as a function of distance from a solar type star at 5 pc to illustrate the gain in detectability when using a 4 QPM coronagraph. *Right:* The detectability of low resolution (R~100) spectra with METIS of giant planets for two ages, 1 Gyr and 0.1 Gyr, as a function of the planet mass.

METIS will detect the faint but characteristic features via eclipse spectroscopy, where spectra during an eclipse are subtracted from spectra shortly before and after the eclipse. Using multiple differential techniques, atmospherics effects can be largely cancelled out, yielding sensitivity to star-planet contrasts well in excess of 10,000. Fig. 3 (right) shows the detectability of low resolution (R~100) spectra of giant planets for ages 1 Gyr and 0.1 Gyr as a function of the planet mass. We use a conservative 1 mJy detection limit, scaled from actual VLT/VISIR observations. On an excellent mid-infrared site, METIS on a 42m E-ELT would be capable of detecting low resolution spectra of a super-Earth orbiting in the habitable zone around one of the stars in the Alpha Centauri triple system, while ocean planets located in the habitable zone could be detected around the Alpha Centauri stars and Barnard's star. For Paranal-like conditions ($m_{M'}$ = 1.2 mag/arcsec to $m_N$ = -5 mag/arcsec), however, the detection of a super-Earth with 2 $R_{Earth}$ located in the habitable zone of either Proxima Centauri or Alpha Centauri A or B would be rather challenging.

## 2.3 Formation and Evolution of Protoplanetary Disks

One of the main surprises in exo-planet research is the large diversity of exo-planetary systems discovered to date, most of which do not resemble at all our own solar system (Udry & Santos 2007). The origin of this diversity must lie in the structure and evolution of the disks out of which they form. The gas in disks plays a key role in the planet formation process, by controlling the dynamics of dust particles and planetesimal growth, providing the main reservoir of material out of which giant planets form, and affecting the migration of planets in the disk (both inwards and outwards). In the most popular core-accretion model, a few rocky cores with masses of 10–20 Earth masses must have formed quickly enough to attract gas to form a gas-rich planet. Over time, at most 20 Myr, the gas in the disk will dissipate and the small grains will coagulate or be blown away. This then leads to the debris disk phase in which the disk is optically thin at UV and IR wavelengths and the grains are of secondary origin, replenished by collisions of larger objects: asteroid-sized bodies or planetesimals.

### 2.3.1 Gas dynamics in planet-forming zones

Gas plays a crucial role in planet formation, not only by providing a necessary ingredient for giant planets but also by affecting the dynamics of the dust particles which coagulate to form planetesimals and eventually terrestrial planets (e.g., Johansen et al. 2007). The CO $v$=1-0 vibration-rotation lines at 4.7μm have proven to be excellent tracers of the warm gas in inner disks as well as the colder gas further out (e.g., Najita et al. 2003). At $R$=100,000 (3 km/s) it is possible to spectrally resolve the line profiles and, through spectro-astrometry, start probing the kinematics down to 0.1 AU scales for a few of the brightest sources (Figure ). METIS with $R$=100,000 will open up the entire variety of disks to such studies, including those in transition to the debris disk phase. The high spectral resolution is essential to probe radii beyond 1 AU, probing departures from Keplerian motion in the disk surfaces and enabling measurements of any radial motion in the gas. For the narrowest case of edge-on disks, the intrinsic turbulent width in the CO emitting layer may be probed, which is connected to the disk viscosity, a quantity intimately related to planet formation.

The main reservoir of the gas in disks is $H_2$, an essential ingredient for building gas giant planets. The pure-rotational mid-IR lines are the lowest possible transitions to search for, but they are extremely difficult to detect because they are intrinsically very weak and always superposed on a strong continuum (e.g., Carmona et al. 2008). METIS can detect ~10$M_{Earth}$ of $H_2$ gas in a disk at 150 pc. For a disk at the distance of the TW Hya association (56 pc), models predict S(2) 12.28μm fluxes around $2\times10^{-15}$ erg s$^{-1}$cm$^{-2}$ if the star has excess UV radiation (Nomura & Millar 2005), readily detectable with METIS.

### 2.3.2 Water and organics in planet-forming zones

Water is one of the most important molecules in planetary systems, not in the least because of its direct association with the biology of living organisms on Earth. Whether or not water is abundant during the formation and early evolution of planets and how the water is transported to the surfaces of terrestrial planets are therefore fundamental questions in planet formation theories. Water is likely the dominant molecule containing oxygen, the third most abundant element in the Universe, thus controlling the chemistry of many other species, and a strong coolant.

An exciting recent development is the detection of strong mid-infrared lines of $H_2O$ and the related OH radical with Spitzer and ground-based 8-meter class telescopes (Salyk et al. 2008, Carr & Najita 2008) (Fig. 4, right). These lines indicate that the water is hot (~800 K) and likely originates from the inner one AU of the disks. However, direct imaging is lacking and line profiles are not fully resolved. Moreover, all detected lines are highly optically thick so that derived abundances have order-of-magnitude uncertainties. Through direct imaging and spectro-astrometry, METIS can determine the location and distribution of water in the inner disk down to 0.1 AU resolution or better and can study its kinematics (R~100,000).

The molecules $CH_4$ (7.7μm), $C_2H_2$ (13.7μm) and HCN (14.0μm) are three key species in the organic chemistry that occurs in the inner (<20 AU) planet-forming zones of disks. Models predict that the abundances of $CH_4$ and $C_2H_2$ peak further away from the star than that of the very stable CO molecule, so their line widths are expected to be narrower, requiring higher spectral resolution. Thus, a second major science goal of this program is to determine the distribution and dynamics of organics in the inner disk. The lines are usually in emission, except for edge-on disks where they will be in absorption. METIS will be able to spectrally resolve the mid-IR lines and to do spatially resolved *absorption* spectroscopy against the disk photosphere.

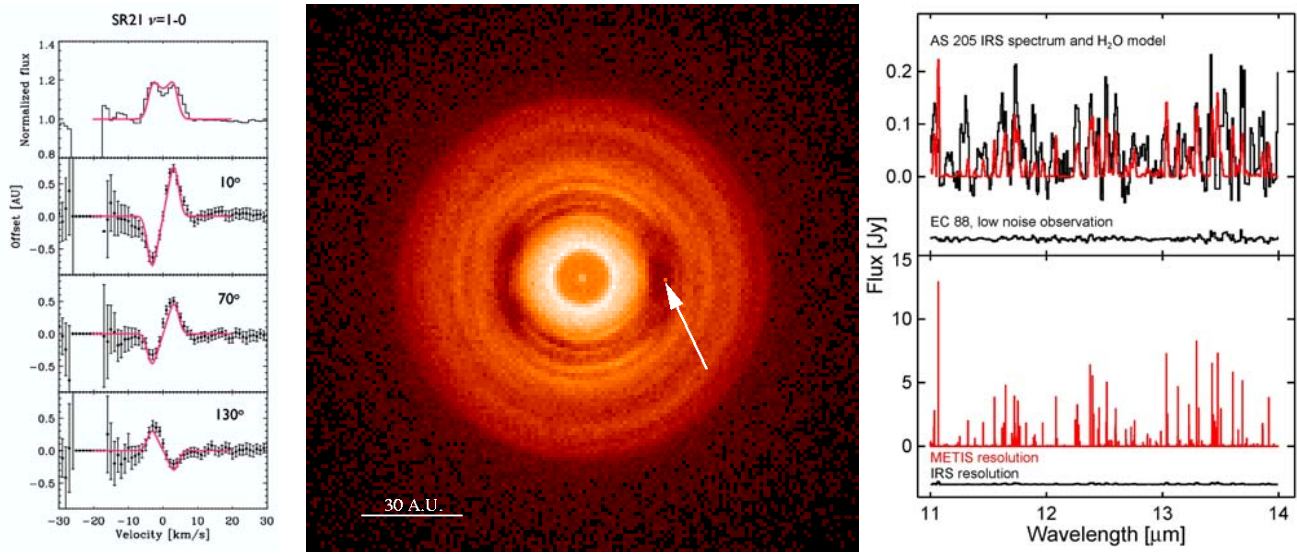

Figure 4 *Left*: The top figure shows an average of CO P-branch lines in the *v*=1-0 fundamental ro-vibrational band at 4.7μm for a proto-planetary disk with a large inner gap, observed with VLT-CRIRES. The lower three figures illustrate spectro-astrometry of these lines for three position angles across the disk; the y-axis shows the offset of the line+continuum signal with respect to the centroid of the continuum emission (Pontoppidan et al. 2008). *Center:* Simulation of the footprints of a 1 $M_{Jup}$ planet, orbiting at 20 AU from a solar type star. The planet clearly leaves its "footprints" on the disk (calculated with the "BUL" code, Charnoz et al. 2001). *Right:* The top figure shows the strong pure rotational lines of $H_2O$ in a Spitzer-IRS $R$=600 spectrum (black) with a model spectrum (red) toward the AS 205 protoplanetary disk. The best fit to the data requires lines from $C_2H_2$, HCN and $CO_2$ in the models. The bottom figure shows the predicted water spectrum at $R$=100,000. Note the greatly increased feature-to-continuum ratio compared to Spitzer-IRS.

### 2.3.3 Dust gaps as signatures of young planets and the exozodiacal dust

Giant planets must form early in the formation of planetary systems, when gas is still present in the disk, typically within 10 Myr after collapse of the cloud that formed the parent star. While such young planets cannot be imaged directly, their presence can be inferred from their interaction with the gas and dust in the disk. An example of such a planet footprint is illustrated in Fig.4 (center). Planets less massive than Saturn will migrate within the disk whereas more massive planets will create a gap in the (dust) disk. However, holes in dust disks can also result from photo-evaporation of the gas and dust, or from grain growth resulting in a lower opacity. Spatially resolved, direct mid-IR images, combined with photometry from other wavelengths, are key to distinguish these possibilities and provide additional constraints on the disk geometry, including flaring.

At later stages, if planets orbit around a star, it is quite plausible that asteroids and comets are also present and then exo-zodiacal dust is produced by asteroid collisions and out-gassing of comets. The observational search for exo-zodiacal radiation is a very active field because this information is needed for any future space missions aimed at direct imaging of terrestrial planets in habitable zones. So far, searches with ISO and Spitzer have been limited to exozodis about 1000 times brighter than the zodiacal light. However, the limitation is not the sensitivity, but the stellar and dust emission have to be spatially disentangled. Typically, an angular resolution of 0.1" (1 AU for a star at 10 pc) is required, corresponding to a telescope diameter greater than 25 meters, when observing at 10μm. The ELT is potentially *the* machine to study exozodiacal light in the 1 AU region around nearby stars (<10 pc). Given the experience with VISIR (detection of the SN1987A ring at a level of about 10 mJy), one can be confident that METIS will have the required sensitivity.

### 2.4 The Growth of Super-massive Black Holes in Galactic Nuclei

In recent years it has become clear that supermassive black holes (SMBHs) form an integral part of galactic nuclei. The finding that central black hole mass and total galaxy mass are closely related is truly remarkable, given that there is a factor of ~$10^8$ between the AU-size Schwarzschild radius of the black hole and the kiloparsec-size dimension of the galaxy. This is evidence that the formation of the black hole is directly related to the formation process of the stellar population, e.g., in a violent burst of star formation. However, the physical process giving rise to the black hole – spheroid mass correlation is not understood at all. Popular scenarios include feedback from an active galactic nucleus (AGN) dispersing the gas feeding both starburst and AGN, and Eddington-limited starburst activity (with radiation pressure from the starburst dispersing the gas). However, none of these scenarios can be tested with current observational techniques. The picture of the simultaneous buildup of a SMBH with an extreme burst of star formation provides a new context for the luminous and ultra-luminous galaxies (LIRGs and ULIRGs), which are major mergers undergoing strong starbursts, and for which it has long been suggested that these are the birthplaces of SMBHs powering active galactic nuclei. Studies of local ULIRGs demonstrate that they plausibly evolve into moderate mass field ellipticals. The giant ellipticals in the local universe would then be the result of similar but much more extreme events at higher redshifts. Indeed, the recently identified population of submillimetre galaxies (SMGs) provides the more luminous high redshift counterpart of the local LIRGs and ULIRGs and may be responsible for the formation of local massive spheroids and for generating QSO activity at high redshift.

Fundamental observational problems are the high levels of obscuration of the active regions where the starburst and the build-up of the black hole take place, and the high spatial resolution required to resolve the region of dynamical influence of the black hole. Both problems can be solved with a mid-IR spectrograph on an ELT, which allows optimal penetration of the dust shroud, as well as superb spatial resolution and sensitivity. METIS at the E-ELT will be able to probe these processes in local AGN and (U)LIRGs by mapping gas flows and measuring dynamic black hole masses.

### 2.4.1 Black hole masses and accretion in nearby obscured AGNs

Determining SMBH masses via dynamical measures is a well-developed subject at shorter wavelengths, where both gas dynamics (assuming circular orbits) and stellar dynamics are used to characterize any dynamically dominant point source. The *radius of influence* of the SMBH is approximately given by

$$\theta_{BH} = 0.03'' \left(\frac{M_{BH}}{10^8 M_O}\right)^{0.5} \left(\frac{100 \text{ Mpc}}{D}\right)$$

Typical AGN and (U)LIRGs in the nearby Universe are Cen A, Circinus, NGC 4945, NGC 1068, NGC 7582, Arp 220, and NGC 6240, with black hole masses between $10^6 - 10^8$ $M_o$. All of them have radii of influence of less than 0.1" (except for Cen A and NGC 7582), which is beyond the capabilities of 8 meter class telescopes but the resolution of the E-ELT will reveal the full Keplerian motion of the gaseous disk. The central regions of these galaxies are highly obscured. In particular NGC4945 is remarkable since variable hard X-ray emission unambiguously reveals the presence of an AGN, yet the obscuration is so large ($A_V > 50$) that no evidence for an AGN is found at optical or even near-infrared wavelengths.

The black hole mass of NGC7582 has been determined (Wold et al. 2006) using mid-infrared spectroscopy of the [NeII] 12.8 μm line at the diffraction limit of VISIR/VLT, demonstrating the promise of mid-infrared spectroscopy for these observations. A mid-infrared spectrograph at an ELT would therefore open up the detailed study of dynamics and gas flow in accreting gas disks in obscured AGNs to the full sample of available AGNs within about 50 Mpc, and no longer be limited to the subset of AGNs that have luminous $H_2O$ maser emission, needed for VLBI measurements. This allows a much more statistically robust approach than hitherto possible.

### 2.4.2 Silicate features as probes of the circum-nuclear region

The central obscuration of AGN produces a pronounced, broad silicate absorption feature at 9.7μm. Jaffe et al. (2004) used the VLTI to investigate the conditions around the nucleus of the nearby AGN NGC1068 on milli-arcsecond scales. Figure 5 (left) shows MIDI spectra of the very centre with maximum VLTI resolution. Not only does the strength of the silicate absorption feature increase with resolution (i.e., probing smaller and smaller regions around the nucleus), but also does the shape of the feature, suggesting the presence of calcium aluminum-silicate ($Ca_2Al_2SiO_7$), a high-temperature species indicating that the dust grains have been processed by the strong radiation field from the AGN.

Spitzer observations of AGN have shown that the silicate feature can also appear in *emission*. This is the case in most type 1 AGN (such as the QSO 3C249.1 – Fig. 5, right), but also in lower ionization AGN. Silicate emission is generally interpreted as the result of a face-on view of the obscuring torus, but may also arise from parts of the narrow-line region directly illuminated by the AGN continuum. We note that the flux density of 3C249.1 (and many other AGN) is about two orders of magnitude less than that of the nearby AGN NGC 1068. METIS on the E-ELT will be able to provide images and spectra of the regions of silicate emission directly associated with the AGN continuum for a much larger sample of objects than hitherto possible.

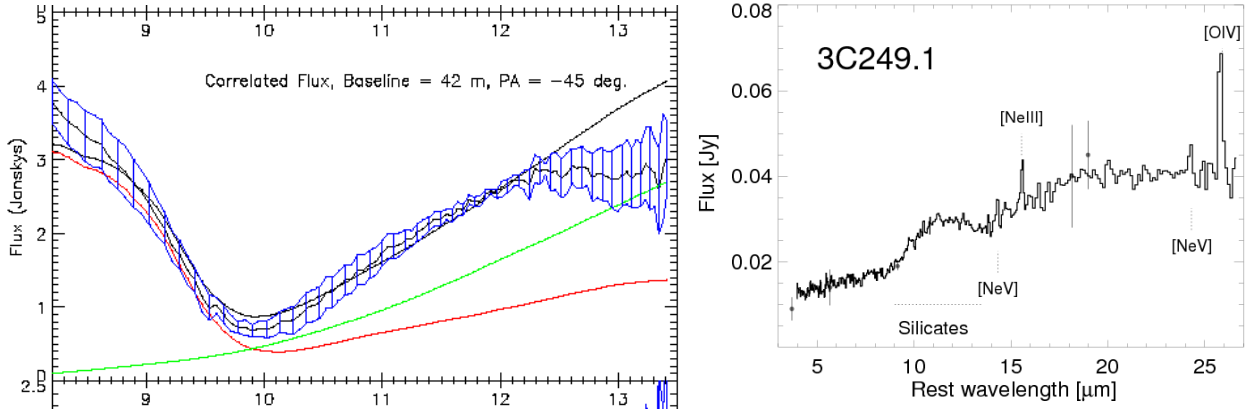

Figure 5 *Left:* VLTI/MIDI N-band spectrum of the very centre of NGC 1068 (Jaffe et al. 2004). *Right:* Spitzer spectra of the QSO 3C249.1, showing the silicate features at 9.7 and 18µm in emission (Siebenmorgen et al. 2005).

### 2.4.3 The starburst – SMBH connection in ULIRGs

ULIRGs harbour the most intense starbursts known in the local universe and are likely the birthplaces of the SMBHs powering AGN activity. They are generally the results of major mergers and in a state of dynamical and structural transformation. As stated above, separating the starburst and AGN contribution in ULIRGs, and measuring the black hole masses, are key problems to be tacked by a mid-infrared ELT instrument. Radio VLBI observations of Arp220 resolve the nuclei into numerous compact point sources, interpreted as supernova remnants. Comparison with the ELT resolution of 0.06" at 10µm shows that the starburst region is easily resolved.

Various mid-infrared spectral features are available to probe different environments in the nuclear region: the ionic lines probe the photoionized gas, PAH emission arises from UV-irradiated photodissociation regions (PDRs) and traces massive star formation, $H_2$ lines probe the warm molecular gas, the broad silicate feature, contains information on the amount of absorbing material along the line of sight and additional information on the AGN. In the vicinity of the AGN significant PAH emission features are usually not observed (e.g., Sturm et al., 2000), because the PAH molecules are destroyed by the hard X-ray radiation. This effect can be used to separate the contributions from starburst and AGN to the total mid-infrared output of the ULIRG, sufficient spatial resolution provided. In addition, the variations in physical conditions can be studied in detail by tracing ionization structure with the [S IV] 10.5µm emission line, the distribution of warm neutral gas using the $H_2$ S(3) 9.66µm line and from the amount and nature of the material obscuring the source by modelling the silicate absorption feature.

The power of this approach lies in the *direct* determination of the origin of the infrared luminosity. Other methods for separating starburst and AGN components (e.g., via [NII]/Hα ratio, presence of broad or high-excitation lines, presence of a non-thermal radio VLBI nucleus, presence of a hard X-ray nucleus) all can be (and have been) used – but not to determine how much the AGN component contributes to the far-IR luminosity. In the mid-infrared, the heated dust and its heating source are probed directly, establishing a direct link with the power source of the far-IR emission.

## 2.5 Science Case Requirements and Comparisons with other Facilities

In addition to the four main science drivers several other areas of great scientific interest, to which METIS is expected to contribute substantially, have been identified (see Brandl et al. (2008) for more details). These additional science cases include:
- The formation of massive stars
- The Galactic Center region
- The life cycle of dust
- The IMF in massive star forming regions
- The hosts of sub-millimeter galaxies
- Gamma-ray bursts as cosmological probes

Fortunately, the instrument requirements from these additional cases are in excellent agreement with the requirements defined by the four main science drivers. Specifically, they require diffraction limited performance, a wavelength coverage from L through N band (3 - 14μm) and both imaging and spectroscopic modes. The cases of the Solar System and protoplanetary disks also require high resolution spectroscopy of R~100,000. Since METIS is more likely to investigate known targets rather than performing deep all-sly surveys, the field of view can be rather small, matched to the isoplanatic angle of a few tens of arcseconds at mid-IR wavelengths. For spectroscopy both long slit and integral field unit (IFU) spectrographs need to be considered, depending on the specific application. A coronagraph will be required for the case of exoplanets.

Here we briefly compare METIS with the abovementioned modes to state-of-the-art facilities that will be available at the time METIS is expected to operate. **ALMA** observations at the highest frequencies and longest baselines would be able to observe AGN at an angular resolution approaching that of METIS but only using molecular gas features. It is however unlikely that these molecules can survive in the region of interest, close to the central engine. In nearby protoplanetary disks, ALMA's sensitivity allows the observations of spectral lines down to 5-10 AU (even for the brightest CO lines) but not to 1 AU or less. Thus, ALMA will focus largely on the kinematics and chemistry in the outer disks (>10 AU), whereas METIS will be unique in the inner disk. Moreover, ALMA cannot observe symmetric molecules without a dipole moment, such as $CH_4$ and $C_2H_2$, which are prime building blocks of organics. ALMA will also be able to contribute to the exploration of comets, but at a very high cost in terms of observing time as the numerous spectral features need to be explored one-by-one, excluding a large sample size for a global study.

For AGN studies the **JWST** would provide excellent sensitivity but does not have the required angular resolution to resolve the region of influence of the black hole. Because of its higher sensitivity JWST-MIRI might be better suited to find cooler, Neptune-like ice giants orbiting nearby stars. However, METIS will be capable of studying giant exoplanets at much smaller separations from their host star, probing a different area in parameter space. The spectral resolution provided by the instruments on JWST is also insufficient to probe kinematics in protoplanetary disks. For solar system observations, METIS will offer a wider range of solar elongation ranges, including opposition observations which are important for thermal modeling aspects (opposition effects, beaming effects, light-curve effects).

METIS will also be competitive to **ELT near-IR** imaging surveys for giant exoplanets around nearby stars because of the higher Strehl ratio at 10μm and the 1000 times smaller contrast between a Jupiter-type planet and a solar-type star. Compared to ELT near-IR observations, which will provide superior angular resolution, METIS will penetrate the obscuring dust in ULIRGs about one order of magnitude better than instruments observing in JHK bands. Similarly, the **VLTI** can reach higher angular resolution than METIS but its sensitivity will only reach the most luminous objects. Finally, **in-situ** measurements by spacecrafts will provide unsurpassed data quality on solar system objects but are limited to a very small number, which will not increase by much over the coming 1 ½ decades (approximately 3 comets and 4 asteroids at best).

## 3. SOME CHALLENGES

The baseline concept of METIS does not require completely new technologies. However, there are numerous challenges that need to be addressed, which goes beyond the scope of this paper. Here we mention just three of them to illustrate specific challenges of mid-IR observations on an ELT.

## 3.1 Background Subtraction Techniques

Ground-based infrared telescopes have to operate at or close to noise limits set by the background radiation from the atmosphere and the thermal radiation from the telescope. The optical surfaces of the telescope and the atmosphere will radiate both continuum and line fluxes in the optical path of any scientific instrument. Cryogenic pupil stops can reduce the straylight but not avoid this temporally variable background. Depending on the detector read noise the shot noise contribution from the thermal background becomes dominant under typical conditions at $\lambda > 2\mu m$. METIS, operating at even longer wavelengths will be background noise limited in all modes, even for high resolution spectroscopy in most parts of the spectrum at integration times of a few seconds. In order to detect astronomical objects that are 5-6 orders of magnitude fainter than the background level careful subtraction and noise filtering techniques need to be applied.

Since the background flux density and structure is time-variable it needs to be measured simultaneously with the signal or very close in time. The former case requires a close-to-perfect flat-field of the detector, and the latter requires "chopping" between source and background-only ("sky") at a frequency of typically a few Hertz for subsequent target-sky subtractions. In contrast to existing infrared telescopes, the E-ELT will not be equipped with a "classical" chopping secondary mirror and other means to switch quickly between target and sky need to be found. In the following we discuss a few alternative techniques to measure and remove the thermal background:

- *Focal Plane Chopping*: Here the focal plane is being moved to alternate between sky and target at typically a few Hertz. This could be achieved in two ways, either by moving periodically the detector itself or by moving a kinematic mirror which moves the image on the detector. The former would be a theoretically very elegant method, but depends on whether such a technically challenging scheme for a large focal plane can be developed and run reliably. The latter has the advantage that the chopping frequency can be high and several chopping distances can be easily achieved (which is difficult with conventional chopping), but has the severe disadvantage that the images are slightly out of focus. Thus, this may only be an option for very small objects and angular offsets.

- *Pupil Plane Chopping*: This approach is a variation of the "classical chopping" but with the "secondary mirror" now internal to the instrument, however looking upstream through different optical paths. It can be done by means of a relay (e.g. an Offner relay) and should be coupled with the AO system to compensate for the tip-tilt error; the wavefront error associated with the tilt of the mirror would need to be compensated by the AO system. A careful design to avoid large strokes on the deformable mirror may be an excellent way to enable pupil plane chopping. However, a fundamental problem is the interaction with active optics: whenever the movable mirror is ahead of the wave-front sensor this scheme needs a "counter-chopper" which may lead to a high degree of complexity or undesirable settling times to achieve high Strehl ratios.

- *Dicke Switching*: Here a stabilized, internal reference source is being used as "sky reference". This method has been used in the past in radio-astronomy and has practically been applied very successfully to IR observations of the Sun (Deming et al. 1986). The method may not offer the highest sensitivity, but it is most likely the only method to exploit a large field of 20-30 arcsec at an ELT without compromising the spatial information resulting from chopping. This method could also be combined with a rigorous flat-field calibration.

- *Jittering*: Here the telescope pointing follows a sequence of small offsets to move the source within the field of view, typically rather slowly at a few tenths of Hertz. This is the "poor man's method" that can be applied if the above methods fail. At the VLT "jittering" is being used for all near-IR imaging observations for best sky-flats and sky-subtraction. It is the default technique at NIR wavelengths but requires extremely good and stable flat-fields, a requirement that needs to be verified with the new generation of mid-IR detectors. The overall experience from mid-IR observations in the US and at ESO is that the resulting frames suffer from fixed pattern noise. However, further improvements of this method for mid-IR wavelengths appear possible.

The trade-off between the various techniques and possible ways of implementation require further study. It is conceivable that there is not one particular "best concept" but – depending on the observation mode and parameters – different methods may be used or combined.

## 3.2 Adaptive Optics

In order to achieve diffraction-limited performance at wavelengths $\geq 3\mu m$ the E-ELT will require adaptive optics (AO) correction of refractive and dispersive atmospheric effects. If these corrections can be successfully applied, an improvement in angular resolution of almost an order of magnitude at 10μm can be expected. As the Kolmogorov turbulence power spectrum drops steeply with increasing wavelength, diffraction limited performance in the mid-IR should be much easier to achieve. However, applying AO corrections to mid-IR wavelengths may bring its own set of problems:

- Atmospheric refraction shifts the apparent position of astronomical objects towards the zenith $z$, with a shift depending on wavelength. Kendrew et al. (2008) calculated a broadening of the PSF of approximately 33 and 11 milliarcsec at L ($\Delta\lambda=0.7\mu m$) and N-band ($\Delta\lambda=1.5\mu m$), respectively for Paranal at a zenith angle of 50 degrees. On a higher site like Cerro Macon the dispersion broadening decreases to 25 and 5 milliarcsec, respectively. For L-band these numbers clearly exceed the width of the diffraction-limited PSF and may require atmospheric dispersion correction when using a narrow slit spectrograph.

- If the AO system uses a wavefront sensor (WFS) at visible wavelengths the centers of the optical and mid-IR images are displaced, and the amount of displacement will change with time (zenith distance). Since the WFS will stabilize the optical source the IR image will gradually trail, broadening the PSF (Roe 2002). Kendrew et al. (2008) derived a displacement of approximately 1 arcsec at $z = 45$ on Paranal if the sensing is carried out in the V-band. The problem will be mitigated when the wavefront sensing is carried out at K-band (2.2 μm), suggesting the inclusion of a K-band WFS within METIS. However, this would preclude the use of laser guide stars for increased sky coverage, at least for diffraction limited performance.

- The presence of water vapour is expected to cause increased differential refraction and atmospheric turbulence. The refractive index of air has contributions from both a slow-varying continuum and strong infrared resonances resulting from absorption by water, $CO_2$ and other atmospheric constituents, giving rise to a complex spectrum that varies with atmospheric temperature, pressure and composition. This makes a simple extrapolation of the visible/NIR wavelengths to longer wavelengths unsuitable for precision modeling and limits the accuracy of systems that sense the wavefront in the optical/near-IR and apply the information to mid-IR wavelengths. The exact amount of this effect and its temporal variations need to be further studied.

### 3.3 The Site Choice for the E-ELT

The performance of METIS does significantly depend on the observatory site. Here, the most relevant site parameter is the altitude, which affects the ambient temperature of the telescope, the effective temperature of the atmosphere, and the amount of precipitable water vapour, which limits transmission and adds to the background emission. We have used radiation modeling software including the data bases for spectral lines and compared the results to recent CRIRES measurements, yielding an excellent agreement between models and data.

At very low humidity (typically above 5000 m) even the strongest absorption features are no longer saturated, so that the observed spectra can be calibrated for telluric absorption. During periods of low atmospheric water vapour, the usually opaque 5-8 μm range becomes accessible to ground-based astronomy. The sensitivity of observations longward of 20μm depends dramatically on the altitude. Fig. 6 shows a relative comparison between typical METIS sensitivities achievable from two potential sites: Cerro Macon and Paranal. The figure clearly indicates that, for the band average, the gain in sensitivity from a higher site is about a factor of two at LM band and about a factor of five at Q band, with the smallest improvement at N band. Please note that the corresponding difference in observing time would be the square of these factors!

Fig. 6 also shows that at certain wavelengths in M and Q band observations will essentially be impossible from the low site, resulting in "lost science". These distinct absorption features contribute substantially to the overall background, similar to the OH lines in the NIR spectral range. The drop of the relative sensitivity toward the band edges also indicates that the bands are wider at higher altitude, which can be crucial for studies of redshifted objects. The sensitivity limits of METIS on a high site are listed in Table 2.

Table 2 METIS point source sensitivities (10-σ, 1hr) for imaging, medium- and high-resolution spectroscopy at the E-ELT on a high site (Cerro Macon).

| 10-σ PSS in 1hr | Imaging (R=5) | R = 3000 (IFU) | R = 100,000 (LM) and 50,000 (N) |
|---|---|---|---|
| L (3.5 μm) | 0.2 μJy | 0.01 mJy | $7 \cdot 10^{-22}$ Wm$^{-2}$ |
| M (4.6 μm) | 1.0 μJy | 0.04 mJy | $2 \cdot 10^{-21}$ Wm$^{-2}$ |
| N (10.5 μm) | 8.0 μJy | 0.30 mJy | $6 \cdot 10^{-21}$ Wm$^{-2}$ |

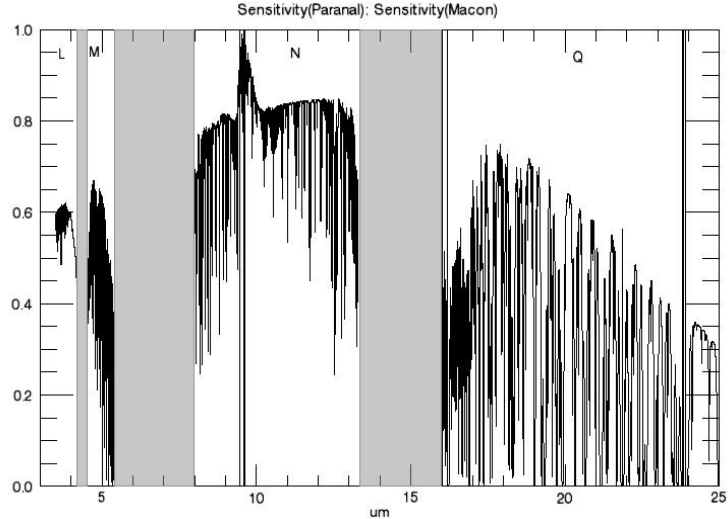

Figure 6 The relative METIS point source sensitivity (Paranal) / sensitivity (Cerro Macon) as a function of wavelength at a resolution of R=3000.

## 4. INSTRUMENT BASELINE

We derived the instrument baseline (below) from the four main science cases for METIS (see section 2). We emphasize that the project is still in phase-A and the concept is not yet fully consolidated.

- Diffraction-limited limited performance at all observing modes
- High-contrast (high dynamic range) imaging at L, M and N band, with fields of view of ~ 20" × 20"
- High resolution (R~100,000) spectroscopy at L, M and N-band
- Coronagraphy
- Low-medium resolution slit spectroscopy

This instrument baseline represents the minimum configuration of an instrument that fully benefits from the outstanding capabilities of the E-ELT. In particular, the close connection between imaging and spectroscopy modes, as emphasized in the science case discussion, is an important characteristic of METIS. In terms of existing instruments at ESO's VLT, the proposed METIS baseline is a combination of the long wavelength channel of CONICA, the long wavelength channel of CRIRES, and the short wavelength channel of VISIR.

In addition to the above baseline configuration several optional instrument modes appear scientifically very attractive. Those include:

- a larger field of view for both imager and IFU spectrograph
- medium resolution IFU spectroscopy
- extended wavelength coverage to include Q-band (site dependent)
- linear polarimetry for both imaging and spectroscopy
- a differential imaging mode
- parallel observing modes.

These optional modes require a careful case-by-case trade-off study between the scientific gain, the technical and operational complexity, and the associated costs and risks.